# Xona Pulsar
# Compatibility with GNSS

Tyler G. R. Reid, Matteo Gala, Mathieu Favreau, Argyris Kriezis, Michael O'Meara,

Andre Pant, Paul Tarantino, Christina Youn

*Xona Space Systems*

**ABSTRACT**

At least ten emerging providers are developing satellite navigation systems for low Earth orbit (LEO). Compatibility with existing GNSS in L-band is critical to their successful deployment and for the larger ecosystem. Xona is deploying Pulsar, a near 260-satellite LEO constellation offering dual L-band navigation services near L1 and L5. Designed for interoperability, Pulsar provides centimeter-level accuracy, resilience, and authentication, while maintaining a format that existing GNSS receivers can support through a firmware update. This study examines Pulsar's compatibility with GPS and Galileo by evaluating $C/N_0$ degradation caused by the introduction of its X1 and X5 signals. Using spectrally compact QPSK modulation, Pulsar minimizes interference despite higher signal power. Theoretical analysis is supported by hardware testing across a range of commercial GNSS receivers in both lab-based simulation and in-orbit live-sky conditions. The study confirms Pulsar causes no adverse interference effects to existing GNSS, supporting coexistence and integration within the global PNT ecosystem.

## 1. INTRODUCTION

More than fifty navigation satellites have been launched in recent years into LEO, with thousands more announced by more than ten emerging providers. These systems range in spectrum usage from traditional RNSS L-bands to C, S, UHF, VHF, and others. Compatibility with existing medium (MEO) and geosynchronous (GSO) orbit GNSS, specifically in L-band, is paramount to the success of their deployment and to navigation users at large. Table 1 summarizes some of these systems.

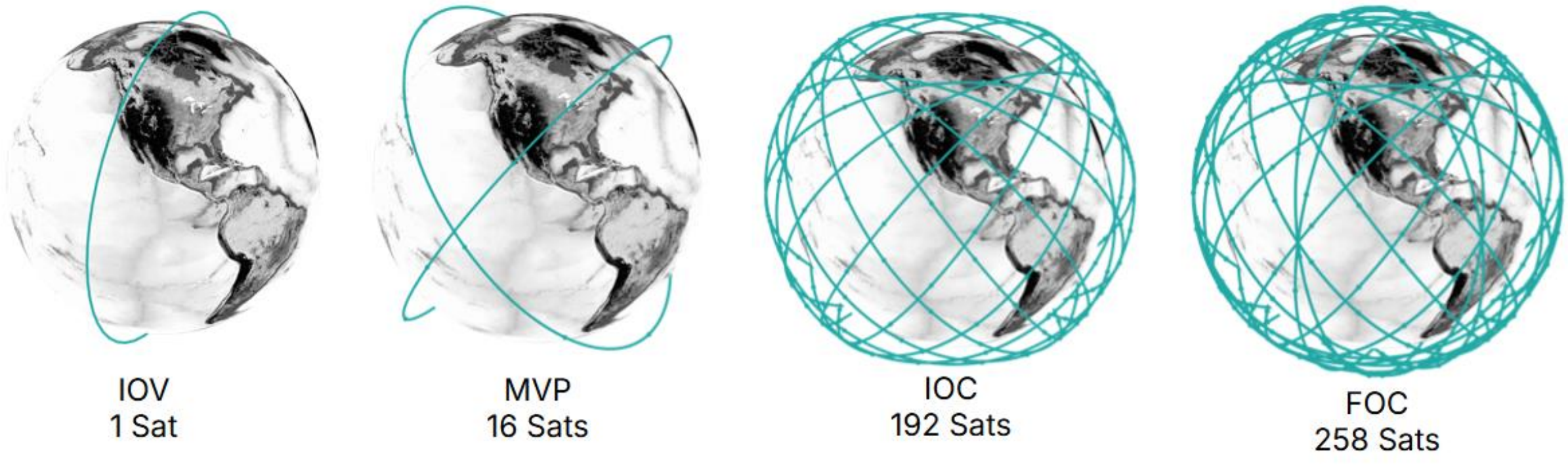

*Figure 1: Xona Pulsar constellation rollout.*

*Table 1: Comparison of dedicated LEO PNT systems, deployments, and plans. Note that satellites already deployed were verified on celestrack.org.*

| Constellation | Flag | Status | Frequency Band(s) | Launches | Constellation Size | Ref. |
|---|---|---|---|---|---|---|
| Iridium STL | USA | Active | L (non-RNSS) | 2017 (NEXT) | 66 (active) | [1] |
| Xona Pulsar | USA | Deploying | L | 2022 (demo)<br>2025 (ops)<br>2026+ (planned) | 2 (launched)<br>258 (planned) | - |
| TrustPoint | USA | Demo / Planned | C | 2023, 2025 (demo)<br>2026+ (planned) | 3 (launched)<br>288 (planned) | [2] [3] |
| ESA LEO-PNT | Europe | Demo / Planned | L, S, C, UHF | 2025-2028 (demo) | 10 (demo, planned)<br>263 (planned) | [4] [5] |
| JAXA | Japan | Feasibility | C | 2030 (P1 planned)<br>2035 (P2 planned) | 240 (P1 planned)<br>480 (P2 planned) | [6] [7] |
| ArkEdge Space | Japan | Feasibility | VHF | - | 50-100 (planned) | [6] |
| Fergani Space | Turkey | Demo / Planned | L, S, Ku, Ka | 2025 (demo)<br>2026+ (planned) | 1 (launched)<br>120 (planned) | [4] [8] |
| CentiSpace | China | Deploying | L | 2018 - 2022 (5 demos)<br>2025+ (10 ops) | 16 (launched)<br>190 (planned) | [6] [9] |
| Geely (Geespace) | China | Deploying | L, S, Ku, Ka | 2022 - 2024 (ops)<br>2025+ (ops) | 25* (launched)<br>240 (planned) | [6] |
| SatNet LEO (Hulianwang) | China | Deploying | L | 2023 – 2025 (ops)<br>2025 (planned)<br>2030 (planned) | 71** (launched)<br>168 (Initial planned)<br>504 (Final planned) | [6] [10] |
| GNSSaS | UAE | Planned | L, S | - | demo planned | [4] |
| VyomIC | India | Planned | - | - | 125 – 150 (planned) | [11] |

*Although 25 Geespace satellites have been launched, it is unclear to the authors if these are part of the dedicated 240 satellites as part of the PNT sub-constellation. The larger Geely constellation has been announced to be 5,676 satellites [6].

**Although 71 Hulianwang satellites have been launched it is unclear to the authors if these are part of the dedicated 508 satellites as part of the PNT sub-constellation. The larger constellation has been announced to be 13,000 satellites [10].

Xona is actively deploying Pulsar, a dedicated constellation of near 260 LEO satellites that will initially offer a dual L-band commercial navigation service adjacent to L1 and L5. Pulsar is designed for GNSS compatibility and interoperability while offering centimeter-level ephemeris, stronger signals for resilience, data encryption, and range authentication for commercial users. Interoperability allows for the simultaneous usage of GNSS and Pulsar for maximum performance. It encourages widespread adoption across multiple industries due to its similarity to existing GNSS signals and has already demonstrated that present receiver architectures can utilize Pulsar with only a firmware update. The constellation rollout is shown in Figure 1.

Here, we showcase compatibility of Pulsar with GPS and Galileo based on Internation Telecommunication Union (ITU) recommendations and extensive hardware testing [12]. Compatibility in this context is measured as interference, specifically degradation of GPS or Galileo due to the introduction of Pulsar signals. We investigate degradation of carrier to noise density ratio ($C/N_0$) of Pulsar X1 on the nearby L1/E1 band as well as Pulsar X5 on the neighboring L5/E5. We present Xona's navigation signal design which highlights a bandwidth efficient form of Quadrature Phase Shift Keying (QPSK). This results in signals familiar in form, function, and bandwidth to existing GNSS, but which roll off quickly in the frequency domain outside of the main lobe, leading to spectral separation. While Xona's total received signal power is significantly higher than GPS and Galileo (up to 20 dB), offering resilience to interference for ground users and service indoors, the combination of center frequency and novel modulation scheme mitigates harmful interference to existing GNSS.

According to ITU recommendations, inter-system interference is assessed through theoretical calculations. However, Xona has gone beyond these guidelines for theoretical calculations by conducting both lab-based and in-orbit testing to evaluate the impact of Pulsar X1 and X5 signals on existing GNSS. Hardware testing was performed using industry standard GNSS simulators modified to generate Pulsar signals along with live-sky Pulsar signals from Xona's In-Orbit-Validation (IOV) satellite, Pulsar-0. A wide range of Commercial-Off-The-Shelf (COTS) GNSS receivers were assessed, representing a broad cross-section of key industry applications including IoT, automotive, agriculture, and heavy industry. Theoretical results show no adverse impact on $C/N_0$ and is validated in the receiver hardware testing campaign.

This work represents a critical step in bridging the gap between existing GNSS and next-generation LEO PNT. The approach presented showcases a powerful methodology for new entrants in the Radio Navigation Satellite System (RNSS) L-band to effectively co-exist alongside established GNSS without causing disruption while facilitating integration into existing receiver platforms. Although RNSS L-band is a precious shared resource and perceived as crowded with established GNSS services, considering new modulations and center frequencies opens the door to new services that promote spectrum efficiency and enhance PNT services for all.

## 2. PULSAR SIGNALS

Two navigation signals are broadcast from Xona Pulsar satellites denoted as X1 and X5 following typical signal naming conventions for GPS and Galileo L1/E1 and L5/E5. The characteristics of the signal are summarized in Table 2 for the Full Operational Capability (FOC) – the full 258 satellite deployment. The technical characteristics reflect Pulsar's regulatory filings and might differ from the specified performance in Xona's Interface Control Document (ICD).

*Table 2: Xona PULSAR signal parameters for FOC.*

| Parameter | X1 | X5 |
|---|---|---|
| Center Frequency (MHz) | 1593.3225 | 1190.51625 |
| 99.5% Bandwidth (MHz) | 1.8 | 17.7 |
| Modulation | EFQPSK | EFQPSK |
| Chip Rate (Mcps) | 1.023 | 10.23 |
| Doppler Range (kHz) | 33.8 | 25.2 |
| Max Single Satellite RIP (dBW) | -138.4 | -136.0 |
| Aggregation Gain Factor (dB)[1] | 13.1 | 9.8 |

[1] Calculated using the methodology outlined in ITU-R M.1831-1 and DO-235C/ DO-292A aviation reference antennas.

Each navigation signal is a composite signal consisting of data (I-channel) and pilot (Q-channel) components that are separately modulated by Pseudorandom Noise (PRN) codes. It is similar in structure to the modernized GPS L5 or Galileo E5a / E5b but incorporates additional features such as bandwidth efficient modulation of the PRN chip sequences.

Navigation data is added to the in-phase (I) channel PRN code (data channel PRN code) and consists of satellite almanac, ephemeris, GNSS corrections, and other data used by receivers in computing time and position. The pilot component is data-less and serves as a pilot signal to provide improved PRN code and carrier phase tracking capabilities. The pilot component is modulated by the pilot-channel PRN code which is also referred to as the Q-channel PRN code, as well as a short, periodic overlay code.

Overlay codes are used to improve tracking performance, facilitate extension of the duration of coherent integration by a receiver, provide a means for data symbol and frame synchronization, and in conjunction with the PRN ranging code on the pilot, synthesize a longer composite code with better autocorrelation and cross-correlation performance.

The bandwidth efficient modulation technique used is Enhanced Feher's Quadrature Phase-Shift Keying (EFQPSK) [13]. It is an enhanced form of a family of XPSK (cross-correlated PSK) signal designs which have good spectral roll-off capabilities and are also resilient to nonlinearities in the satellite transmitter as it is constant envelope. This technique is applied to modulate the data and pilot PRN code chip streams to form a spread-spectrum ranging signal with limited out-of-band spectral content. To be clear, this waveform is not generated through filtering but via the modulation, generated through digital signal processing. Figure 2 shows the normalized Power Spectral Density (PSD) of X1 and X5. Figure 3 shows the time-domain I and Q components of the transmitted X5 waveform in comparison to a more traditionally used Binary Phase Shift Keying (BPSK).

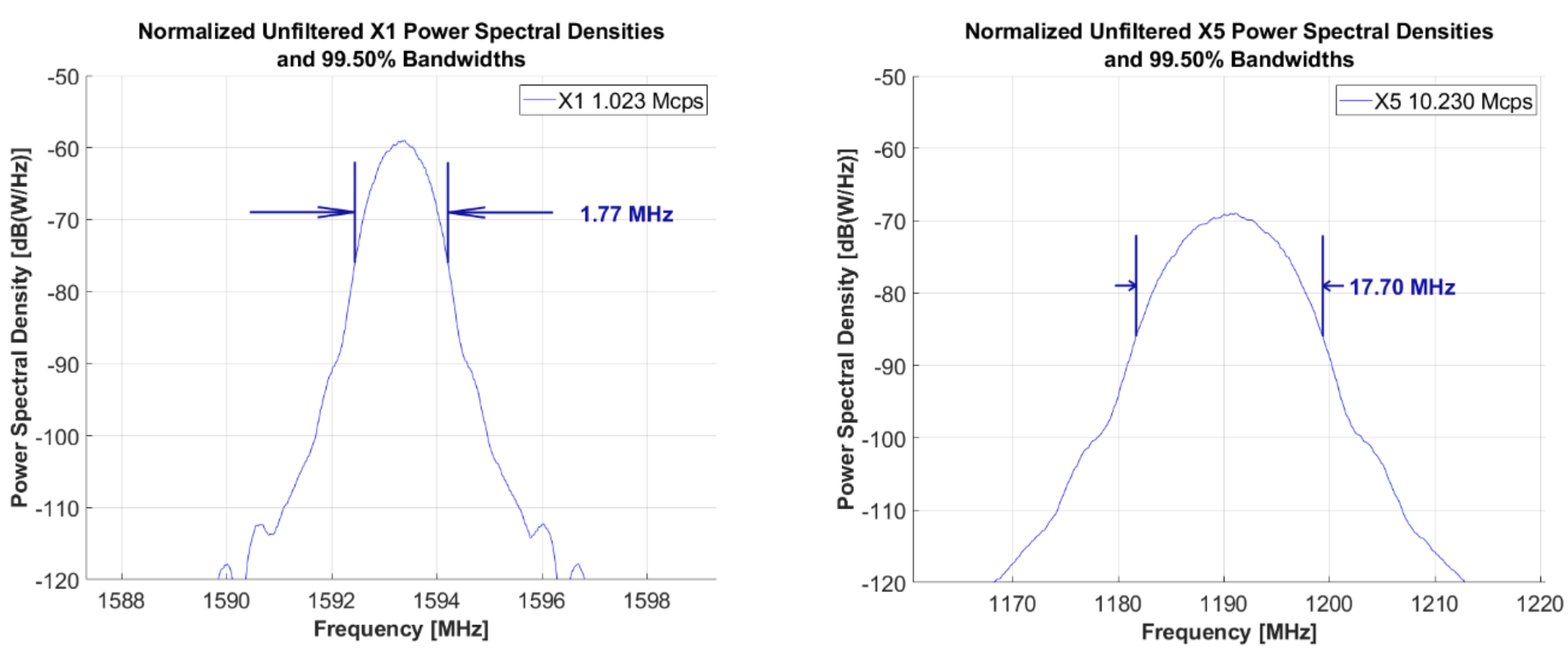

*Figure 2: Normalized Power Spectral Density (PSD) of the Pulsar X1 (left) and X5 (right).*

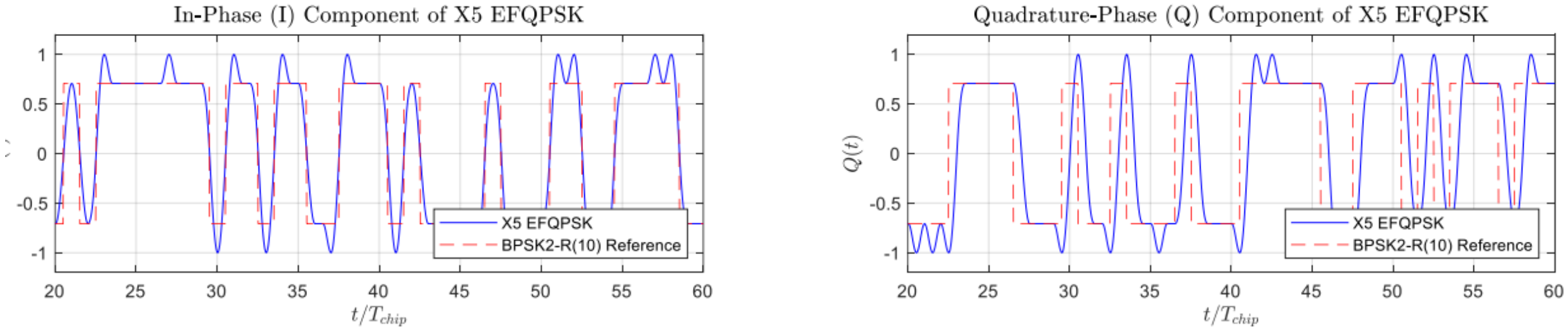

*Figure 3: Pulsar X5 I (left) and Q (right) components.*

Of note is the higher received signal power of the Pulsar signals compared to GPS. In the NAVSTAR GPS Space Segment/Navigation User Interfaces (IS-GPS-200M), the GPS L1C/A received power is stated to be in the range of -158.5 dBW and -153.0 dBW [14]. Table 2 shows the expected maximum received signal power for X1 and X5 as -138.4 dBW and -136.0 dBW, respectively. This is 15 to 20 dB more signal power. Although Xona is adjacent to L1 and L5, the higher signal power is metered by the modulation scheme which limits out-of-band power and hence interference with existing GNSS.

## 3. THEORETICAL COMPATIBILITY ANALYSIS

This section evaluates the interference caused by Pulsar on GPS and Galileo analytically. The metric under consideration is the carrier to noise density ratio, $C/N_0$ and the degradation analysis that evaluates the RNSS inter-system interference. There is no official limit for the maximum $C/N_0$ degradation an RNSS system may cause to incumbent RNSS systems, as the acceptable level of interference is defined during operator-to-operator coordination[1]. ITU document M.1831-1 outlines an analytical approach to $C/N_0$ degradation evaluations for the purposes of inter-system RNSS coordination, which will be the basis for the approach here.

### *3.1 Methodology*

According to ITU M.1831-1 the $C/N_0$ degradation caused by introducing a new radio-navigation satellite system is evaluated by calculating the signal noise environment before and after the new signal. As presented in Equation 1, before the new signal is introduced, there are four types of noise sources, the ambient thermal noise ($N_o$), the noise by adjacent-band non-RNSS system ($I_{ext}$), the noise of the RNSS system to itself ($I_{ref}$) and the noise of other RNSS system currently in operation ($I_{rem}$).

$$\frac{C}{N'_\mathrm{o}} = \frac{C}{N_\mathrm{o} + I_{ext} + I_{ref} + I_{rem}} \tag{1}$$

where:

| | |
|---|---|
| $C$ | Carrier strength of the impacted RNSS signals (W) |
| $N_0$ | Receiver thermal noise power spectral density (W/Hz) |
| $I_{ext}$ | Effective white noise power spectral density (W/Hz) due to non-RNSS “external” signal interference |
| $I_{ref}$ | Effective white noise power spectral density (W/Hz) due to interference from all the signals of the impacted “reference” RNSS systems to itself |
| $I_{rem}$ | Effective white noise power spectral density (W/Hz) due to interference from all existing “remaining” RNSS systems |

Equation (2) presents the $C/N_0$ calculation after the new signal is introduced. $I_{alt}$ represents the added noise from the new system evaluated.

$$\frac{C}{N'_\mathrm{o}} = \frac{C}{N_\mathrm{o} + I_{ext} + I_{ref} + I_{rem} + I_{alt}} \tag{2}$$

[1] For interference from non-RNSS systems to RNSS systems, up to 1 dB degradation is considered acceptable [15].

where:

$I_{alt}$ Effective white noise power spectral density (W/Hz) due to interference from the new "alternate" RNSS system being introduced

ITU M.1831-1 provides reference values for the $N_0$ and $I_{ext}$, with $I_{ref}$, $I_{rem}$ and $I_{alt}$ being calculated according to Equation (3). The effective white noise spectral density ($I$) calculation is a factor of signal strength ($P_{max}^{R}$), the number of satellites ($G^{agg}$), and the Spectral Separation Coefficient ($SSC$) between the signals.

$$I_{ref/rem/alt} = P_{max}^{R} + G^{agg} - L_{proc} + SSC \quad (3)$$

where:

$P_{max}^{R}$ Maximum single satellite received power (W)

$G^{agg}$ Aggregation gain factor for single worst-case satellite geometry (dB)

$L_{proc}$ Processing loss

$SSC$ Spectral Separation Coefficient (Between the impacted and impacting signals)

The $C/N_0$ degradation caused by the introduction of a new system can be simplified as seen in Equation (4), with the resulting equation comparing the noise level added by the new system to the pre-existing noise level. It is important to note that the impact of a new system depends on the level of pre-existing noise.

$$\Delta\left(\frac{C}{N'_{\text{o}}}\right) = \frac{\dfrac{C}{N_{\text{o}} + I_{ext} + I_{ref} + I_{rem}}}{\dfrac{C}{N_{\text{o}} + I_{ext} + I_{ref} + I_{rem} + I_{alt}}} = 1 + \frac{I_{alt}}{N_{\text{o}} + I_{ext} + I_{ref} + I_{rem}} \quad (4)$$

The Xona Pulsar system features stronger RNSS signals than current systems, however, the $C/N_0$ degradation is limited by selecting unused parts of the spectrum and using a bandwidth efficient modulation scheme. In particular, the X1 signal is centered 17.9 MHz from GPS L1 and the X5 signal is centered 14.1 MHz from GPS L5. Xona Pulsar's EFQPSK modulation minimizes the X1 and X5 signal's out-of-band emissions resulting in minimal added noise $I_{alt}$.

### *3.2 C/N$_o$ Degradation Calculations*

In this study, the analytical $C/N_0$ degradation calculations are performed for the GPS L1, Galileo E1, GPS L5, and Galileo E5 signals according to the system parameters published in ITU-R M.1787-5. The same methodology can be applied for evaluating the $C/N_0$ degradation to all systems published in ITU-R M.1787-5 but is not shown here. According to the ITU methodology, all inputs assume a 0 dBi receiver antenna, and the aggregate gain factor is calculated using the single worst-case maximum with an aviation reference antenna according to DO-235C/ DO-292A.

Table 3 presents the calculation inputs from all the values not related to the Xona Pulsar system. The inputs are sourced from ITU-R M.1787-5 and the interference noise calculations are made according to ITU-R M.1831. The systems used for the $I_{rem}$ calculations were selected based on the available information from ITU-R M.1787-5. Some systems with earlier ITU filing dates were not included because their data have not been published in M.1787 to date. As a result, the calculated $C/N_0$ degradation is more conservative than in real-life, as the actual noise environment prior to the introduction of Xona Pulsar would be higher when accounting for those excluded systems.

*Table 3: C/Nₒ degradation calculation non-Xona inputs.*

| Parameter | Values | | | | Notes |
|---|---|---|---|---|---|
| | GPS | GAL | GPS | GAL | |
| | L1 C/A | E1A | L5 | E5 | |
| $N_o$ [dB(W/Hz)] | -201.50 | | | | According to ITU-R M.1831 |
| $I_{ext}$ [dB(W/Hz)] | -206.50 | | | | According to ITU-R M.1831 |
| $L_{proc}$ [dB] | 1.0 | | | | According to ITU-R M.1831 |
| $I_{ref}$ [dB(W/Hz)] | -205.81 | -215.34 | -213.93 | -216.60 | Using Equation 3 and ITU-R M.1787-5 inputs |
| $I_{rem}$ [dB(W/Hz)] | -204.71 | -214.77 | -208.99 | -210.86 | Using Equation 3 and ITU-R M.1787-5 inputs |
| Systems in $I_{rem}$ calculation | GPS/Galileo, GLONASS, BeiDou, QZSS, NavIC/GAGAN, SBAS | | | | According to ITU-R M.1787-5. |
| $N_o+I_{ext}+I_{ref}+I_{rem}$ [dB(W/Hz)] | -198.15 | -200.02 | -199.59 | -199.85 | Total pre-existing noise |

The $I_{alt}$ calculations of the effective white noise introduced by the Pulsar X1 and X5 signals are performed using Equation (3) and the inputs from Received Isotropic Power (RIP) in Table 2. While the transmission parameters of each signal remain constant, the resulting $I_{alt}$ for each existing GNSS signal is different due to the Spectral Separation Coefficient (*SSC*). Table 4 presents the *SSC*, $I_{alt}$ and resulting *C/N₀* degradation for each Xona signal and the evaluated existing GNSS signals.

*Table 4: Analytical C/Nₒ degradation results.*

| Parameter | Values | | | | Notes |
|---|---|---|---|---|---|
| | GPS | GAL | GPS | GAL | |
| | L1 C/A | E1A | L5 | E5 | |
| X1 | | | | | |
| SSC [dB/Hz] | -97.41 | -85.23 | - | - | X1 and existing GNSS signal |
| $I_{alt}$ [dB(W/Hz)] | -223.71 | -211.53 | - | - | Using Equation (3) |
| *C/Nₒ* Degradation [dB] | -0.01 | -0.30 | - | - | |
| X5 | | | | | |
| SSC [dB/Hz] | - | - | -85.04 | -85.99 | X5 and existing GNSS signal |
| $I_{alt}$ [dB(W/Hz)] | - | - | -212.24 | -213.19 | Using Equation (3) |
| *C/Nₒ* Degradation [dB] | - | - | -0.23 | -0.20 | |

For the scope of this paper, two GPS and two Galileo signals are presented analytically, however, similar analysis has been performed for all current GNSS systems. For completeness of the GPS and Galileo analysis, Table 5 presents the resulting *C/N₀* degradation for all signals, showing minimal degradation.

*Table 5: GPS and Galileo C/Nₒ theoretical degradation results.*

| GNSS Signal | X1 | | GNSS Signal | X5 | |
|---|---|---|---|---|---|
| | SSC | C/Nₒ Degradation | | SSC | C/Nₒ Degradation |
| GPS L1-C/A | -97.41 | -0.01 | GPS L2-PY | -93.67 | -0.04 |
| GPS L1-PY | -87.86 | -0.15 | GPS L2-M | -90.56 | -0.08 |
| GPS L1-M | -106.75 | -0.00 | GPS L5 | -85.04 | -0.24 |
| GPS L1C | -86.43 | -0.17 | WAAS L5 | -85.04 | -0.24 |
| WAAS L1 | -97.41 | -0.01 | GAL E5 | -85.99 | -0.21 |
| GAL E1-A | -85.23 | -0.30 | GAL E6-BC | -104.77 | -0.00 |
| GAL E1-BC | -86.43 | -0.17 | GAL E6-A | -96.16 | -0.02 |

## 4. LAB HARDWARE TESTING

ITU recommendations rely exclusively on theoretical $C/N_0$ degradation calculations to evaluate the compatibility of new systems with existing ones. Xona has gone beyond these baseline assessments to ensure that its Pulsar system does not introduce harmful interference to existing systems operating in adjacent frequency bands by including hardware testing both in the lab and with live satellite signals.

Xona leveraged GNSS simulators and COTS receivers for lab-based testing. While theoretical $C/N_0$ degradation estimates are based on added noise from the new system, real-world impacts also depend on receiver-specific signal processing. Each GNSS receiver employs unique radio frequency (RF) front ends and software algorithms, which can significantly influence observed degradation.

To ensure a comprehensive and representative compatibility evaluation, Xona selected a diverse set of receivers spanning a range of price points and use cases. This included three survey-grade, two automotive, and two IoT receivers from several vendors – all COTS devices on the market today. More survey-grade receivers were used as they were found to have the largest variability in results. The aim is to showcase representative behaviour, not single out specific vendors, which is why we will refer to these receivers with a generic naming convention.

### *4.1 Hardware in the Loop Testing*

The Hardware-in-the-Loop (HIL) test setup was designed to emulate the methodology outlined in ITU-R M.1831-1, which is used in the theoretical $C/N_0$ degradation calculations. The test incorporates three key RF components: a simulated reference constellation with GPS L1/L5 and Galileo E1/E5, Pulsar X1 or X5 aggregate power signals, and a noise source that includes thermal noise ($N_0$), external interference ($I_{ext}$), reference interference ($I_{ref}$), and remaining interference ($I_{rem}$). To ensure consistent and repeatable results, Pulsar satellite signals are maintained at fixed positions with constant transmit power. The $C/N_0$ values of GPS and Galileo are measured before and after the introduction of the Pulsar signals.

Figure 4 shows the hardware simulator setups. Xona repeated the experiment on both Safran (Skydel) GSG-8 and Spirent PNT-X simulators to ensure consistency of results. Each was connected to all receivers under test simultaneously conductively via an RF splitter as shown in Figure 5. Figure 6 shows an example of the signals in the frequency domain used in this testing.

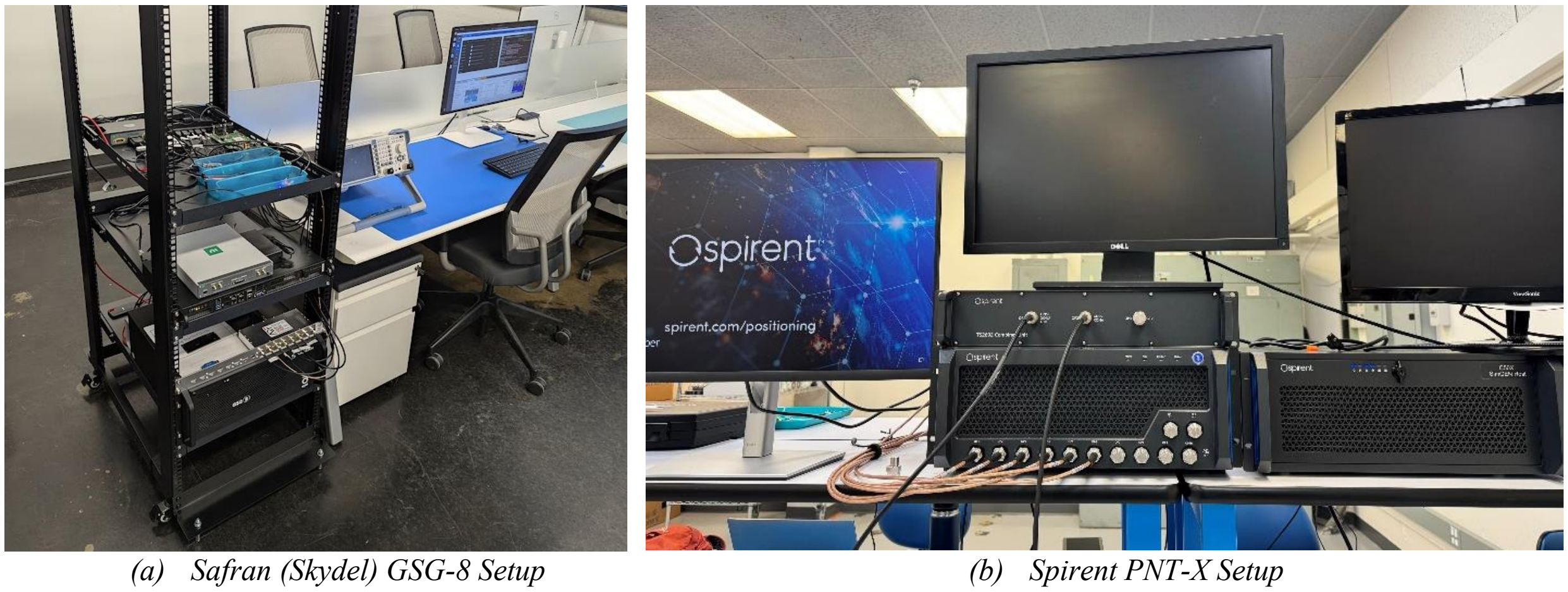

*(a) Safran (Skydel) GSG-8 Setup* *(b) Spirent PNT-X Setup*

*Figure 4: Hardware-In-The-Loop set ups used in hardware receiver compatibility testing. Both setups can generate Pulsar X1 / X5 signals in addition the GPS L1 / L5 and Galileo E1 / E5.*

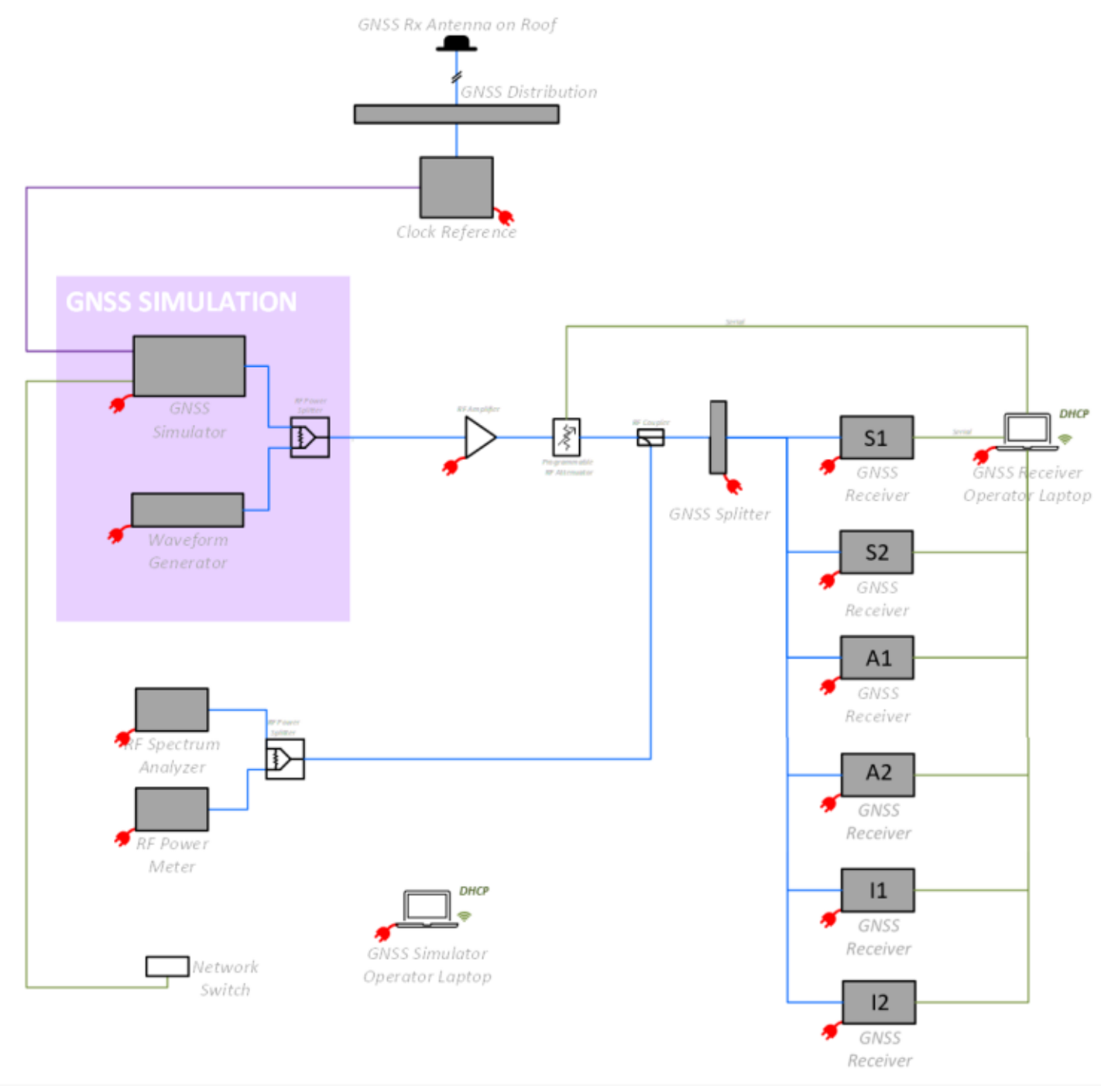

*Figure 5: Experimental set up for conductive compatibility testing.*

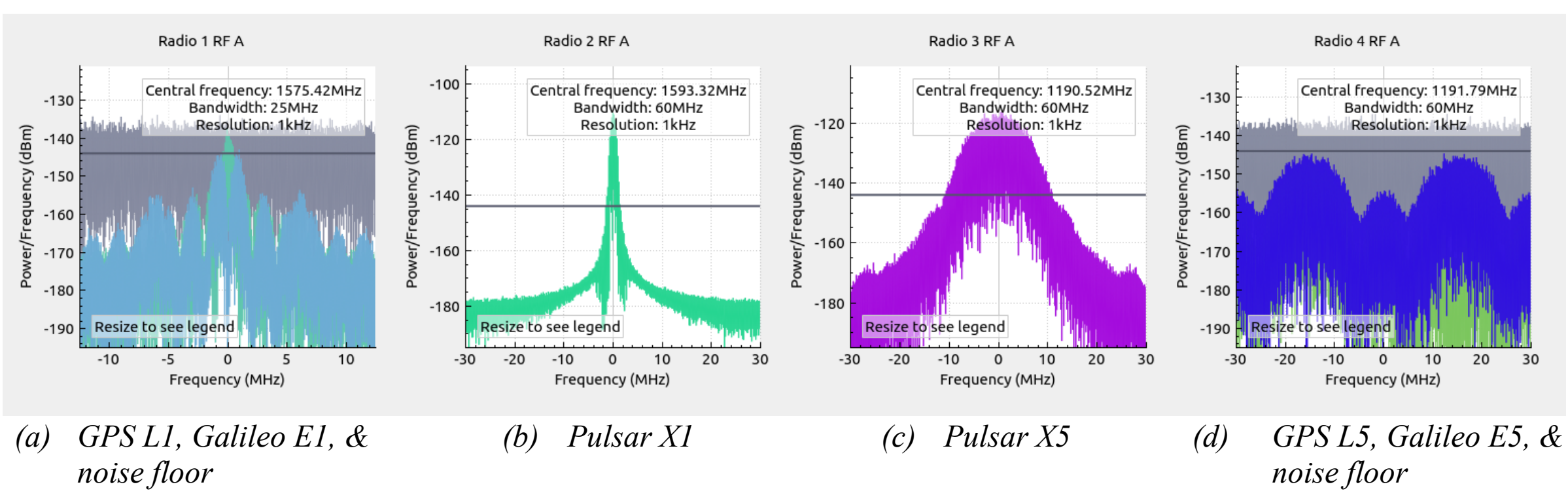

*(a) GPS L1, Galileo E1, & noise floor* *(b) Pulsar X1* *(c) Pulsar X5* *(d) GPS L5, Galileo E5, & noise floor*

*Figure 6: Example spectrum used in conductive hardware testing (from Safran GSG-8).*

For the GPS and Galileo signals, the minimum received power levels specified in ITU-R M.1787-5 were used. For Pulsar signals, the aggregate received power levels were based on the values provided in Table 2. A fixed noise floor of –200.3 dB(W/Hz) was applied across all test cases to simplify the evaluation process. This value represents a conservative assumption, as it is lower than the noise floor calculated in the theoretical compatibility assessment. The selected input parameters are representative of the Pulsar Full Operational Capability (FOC) system as received by a GNSS receiver with a 0 dBi antenna gain.

The experiment was conducted as follows:

(1) All six receivers were connected to the simulator RF output via a splitter.
(2) Receivers were power cycled and cold started, clearing any previous data in memory.
(3) Data logging for each receiver was commenced.
(4) The simulator scenario was commenced. The scenario runs as follows:
    a. GPS + Galileo with noise floor for 10 minutes, allowing receivers to stabilize.

b. Ramp up Xona signal power at 5-minute intervals at the following profile:
   i. Current Antenna Design Levels (CADL), as is on orbit for Pulsar-0.
   ii. Pulsar-0 (IOV) power levels as filed with the ITU.
   iii. FOC aggregate power levels as filed with the ITU.
   iv. Higher power levels (FOC+5dB, +10dB, +20dB, +30dB) to assess sensitivity and margin.
c. Back to GPS + Galileo with noise floor only to assess hysteresis.

This scenario was run over 100 times to assess repeatability and gather statistics. Typical response profiles for a survey receiver (in this case S2) are shown in Figure 7 and Figure 8 for X1 and X5, respectively. It should be noted that Xona does not intend to operate at power levels beyond those filed with the ITU. Higher levels are shown to assess sensitivity only.

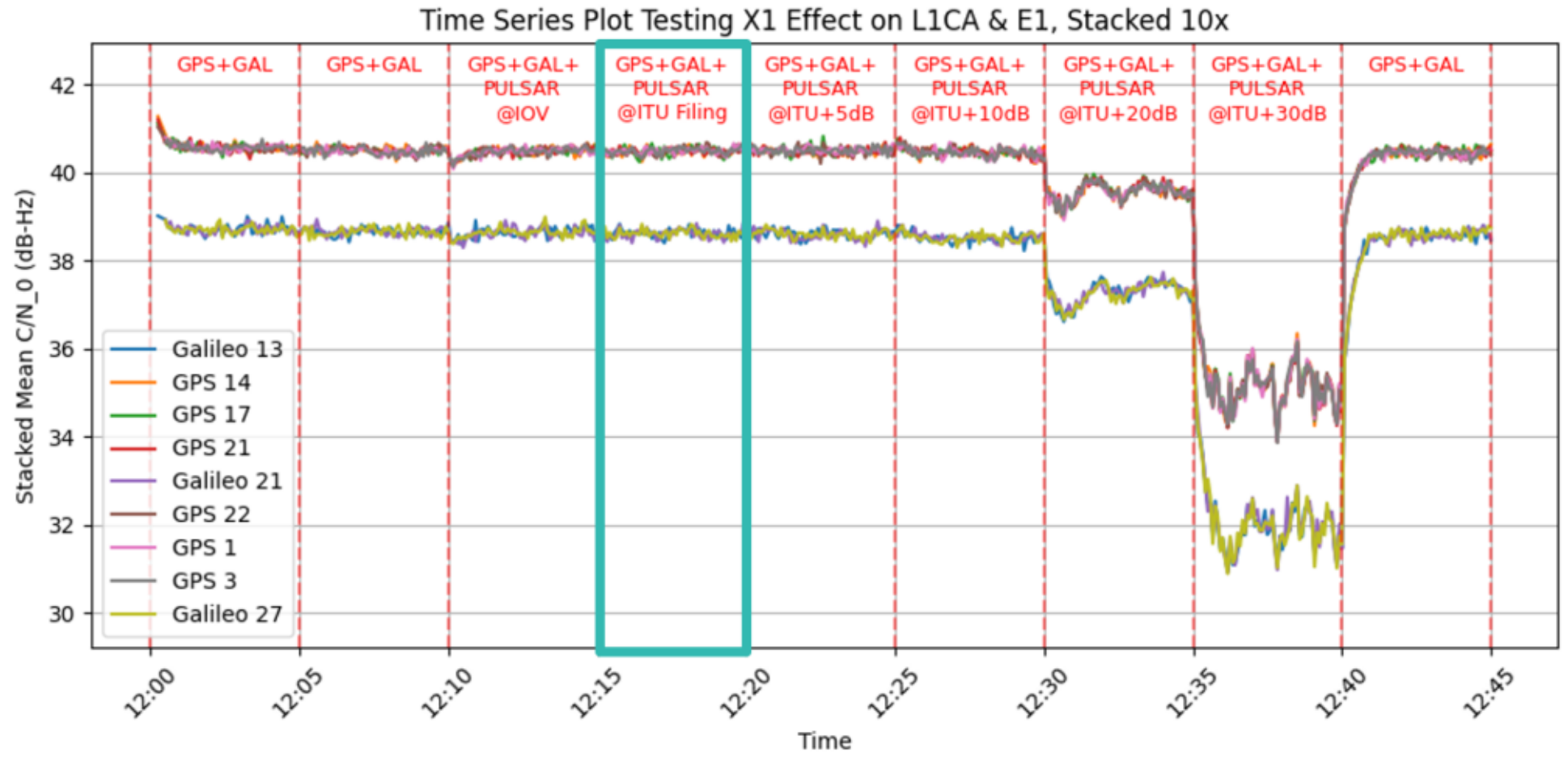

*Figure 7: $C/N_0$ response profile for a typical survey receiver under varying Pulsar X1 power levels for GPS L1 and Galileo E1. This shows both ITU filed (turquoise box) levels as well as higher powers to assess sensitivity.*

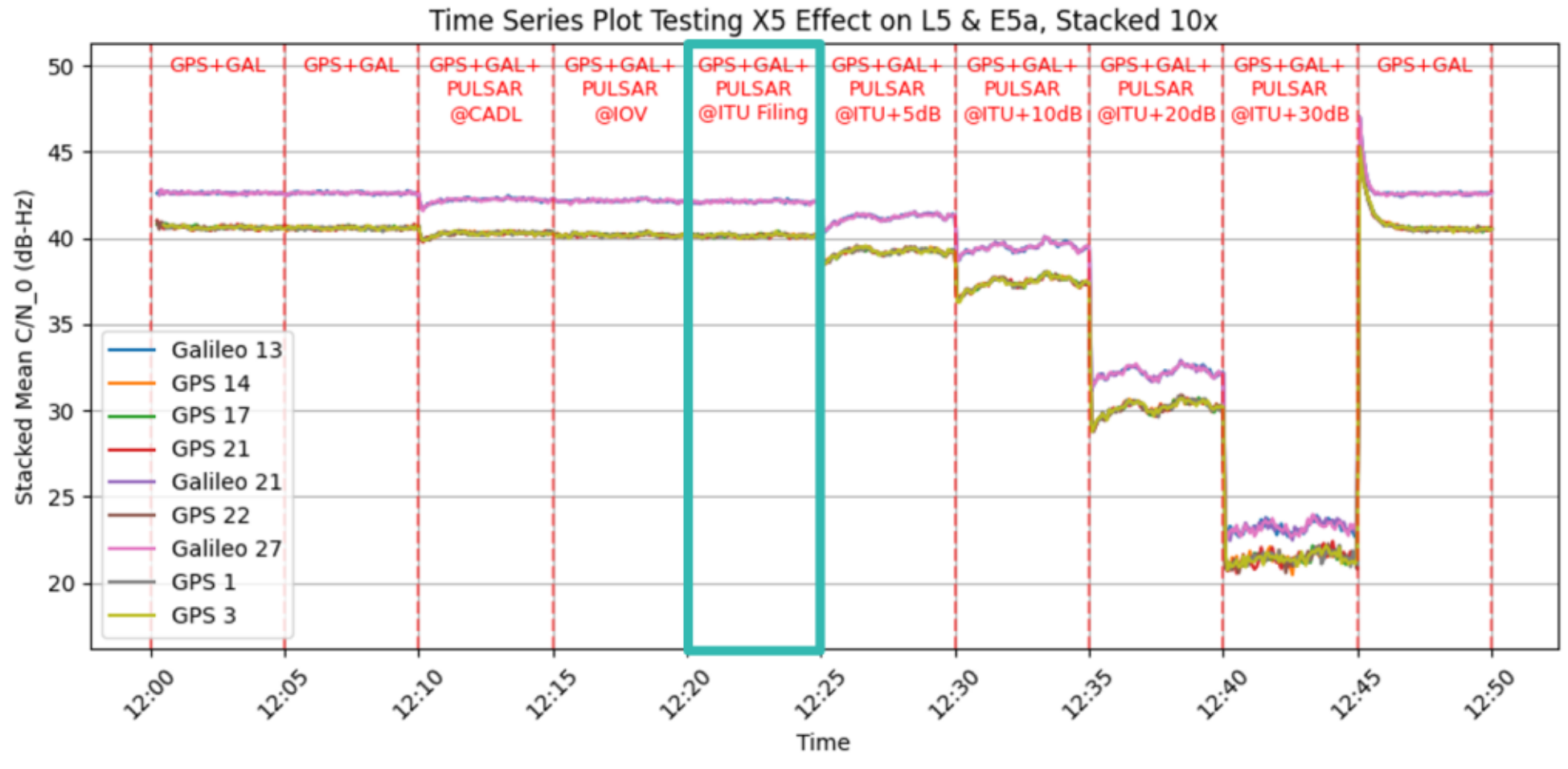

*Figure 8: $C/N_0$ response profile for a typical survey receiver under varying Pulsar X5 power levels for GPS L5 and Galileo E5. This shows both ITU filed (turquoise box) levels as well as higher powers to assess sensitivity.*

Table 6 presents the results of conductive receiver testing, including individual receiver outcomes for the seven under consideration. This includes errors bounds derived from the statistics of the more than 100 runs completed. The HIL conductive testing results highlight the varying responses of individual receivers to the introduction of Pulsar signals at FOC aggregate constellation levels. Overall, the average $C/N_0$ degradation observed across all receivers is consistent in magnitude with the theoretical predictions, lending validation to the theoretical compatibility assessment methodology.

*Table 6: Conductive testing $C/N_0$ degradation results.*

| Xona ITU Filing Parameters | | $C/N_0$ Degradation [dB] | | | | | | | | |
|---|---|---|---|---|---|---|---|---|---|---|
| | | GNSS Signal | Theory | Survey | | | Automotive | | IoT | |
| Pulsar Signal | Aggregate Power [dB] | | | S1 | S2 | S3 | A1 | A2 | I1 | I2 |
| X1 | -125.3 | GPS L1C/A | -0.01 | -0.14 ± 0.01 | -0.05 ± 0.02 | -0.01 ± 0.02 | -0.002 ± 0.00 | -0.01 ± 0.03 | 0.00 ± 0.00 | 0.00 ± 0.00 |
| | | GAL E1BC | -0.17 | -0.17 ± 0.08 | -0.06 ± 0.03 | - | -0.07 ± 0.01 | -0.03 ± 0.00 | 0.00 ± 0.00 | - |
| X5 | -126.0 | GPS L5 | -0.24 | -0.58 ± 0.01 | -0.45 ± 0.02 | -0.38 ± 0.03 | - | -0.22 ± 0.07 | - | - |
| | | GAL E5 | -0.21 | -0.83 ± 0.02 | -0.51 ± 0.02 | - | - | - | - | - |
| | | GAL E5a | -0.30 | -0.54 ± 0.01 | -0.49 ± 0.03 | - | - | -0.33 ± 0.01 | - | - |
| | | GAL E5b | -0.20 | -0.16 ± 0.03 | -0.43 ± 0.02 | - | -0.001 ± 0.00 | - | - | - |

*Entries marked with a '-' indicate the receiver was not capable of tracking that signal.

### *4.2 Anechoic Chamber Testing*

While conductive testing offers precise control by feeding simulated signals directly into the receiver, it may not fully capture real-world signal propagation effects. To address this limitation, additional receivers were also evaluated in an anechoic chamber. The testing methodology mirrored that of the conductive setup, using the same three RF components: simulated GPS + Galileo signals, Pulsar X1 + X5 FOC aggregate power signals, and noise floor. Transmit power levels were adjusted to account for free-space propagation losses within the chamber, ensuring that the effective Received Input Power (RIP) at the receiver antenna matched that used in the conductive tests. Figure 9 presents the hardware configuration for the chamber test setup located at Xona facilities.

Four survey-grade receivers were tested in this set up. S1 was used in both chamber and conductive testing to serve as a common element for comparison. Three additional survey-typer receivers, denoted S4-S6, were used due to both the observed variability in survey receivers in conductive testing along with availability of certain hardware during testing. Table 7 summarizes these results. In some cases, receivers even exhibited an increase in $C/N_0$ values. While this is not physically possible in terms of pure signal-to-noise physics, it can be attributed to internal signal processing behaviours unique to each receiver. The anechoic chamber results further validate the theoretical calculations methodology and demonstrate Pulsar's compatibility in lab conditions that are closest to the real world.

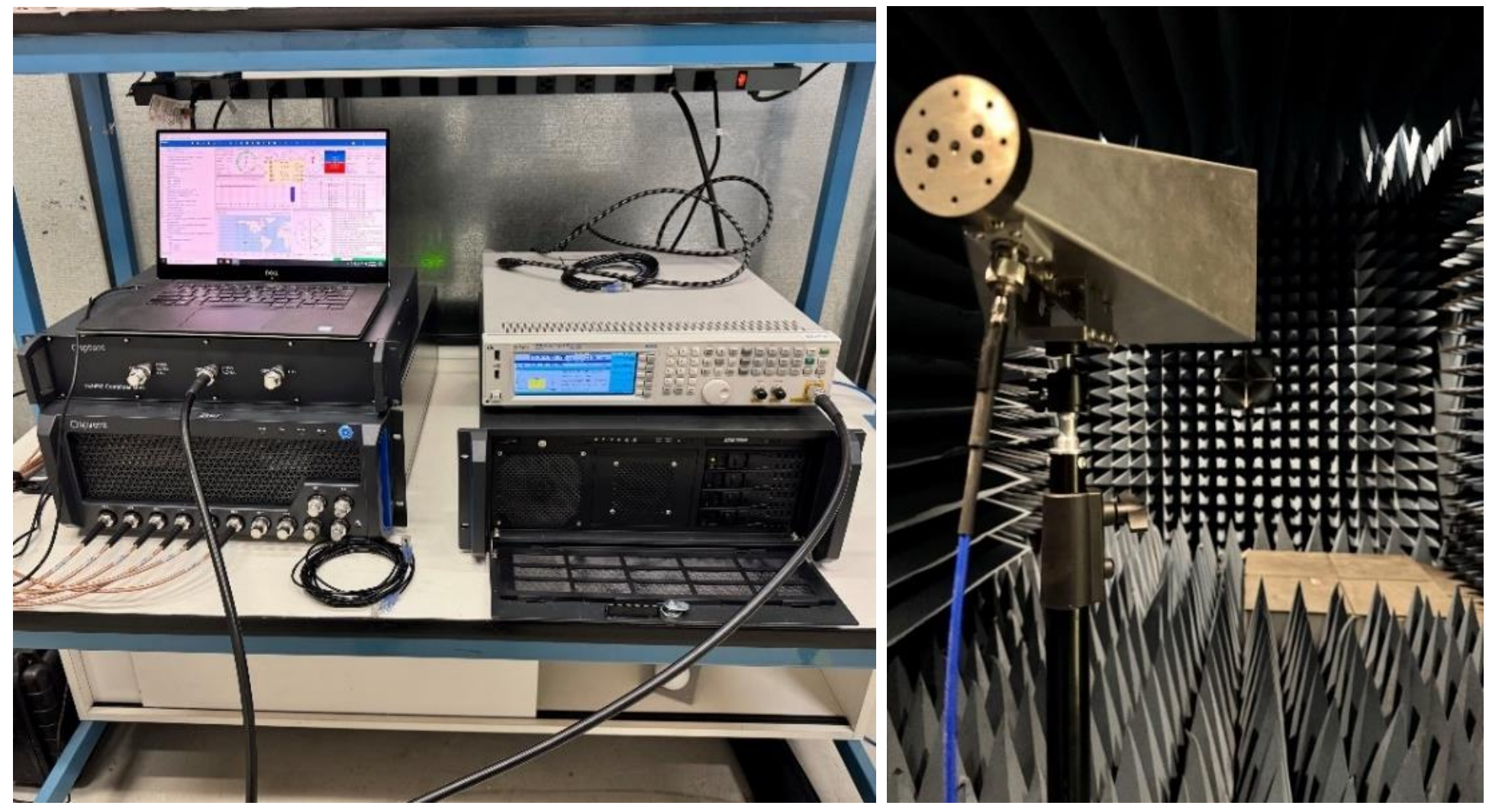

*Figure 9: Hardware test setup for anechoic chamber compatibility testing.*

*Table 7: Anechoic chamber $C/N_0$ degradation results.*

<table>
<tr><th colspan="2">Xona ITU Filing Parameters</th><th colspan="2" rowspan="2">GNSS Signal</th><th colspan="6">$C/N_0$ Degradation [dB]</th></tr>
<tr><th>Pulsar Signal</th><th>Aggregate Power [dB]</th><th>Theory</th><th>Receiver Average</th><th>S1</th><th>S4</th><th>S5</th><th>S6</th></tr>
<tr><td rowspan="2">X1</td><td rowspan="2">-125.3</td><td colspan="2">GPS L1-C/A</td><td>-0.01</td><td>+0.10</td><td>+0.17</td><td>0.00</td><td>-0.01</td><td>+0.23</td></tr>
<tr><td colspan="2">GAL E1-BC</td><td>-0.17</td><td>+0.11</td><td>+0.15</td><td>-</td><td>-0.01</td><td>+0.19</td></tr>
<tr><td rowspan="3">X5</td><td rowspan="3">-126.0</td><td colspan="2">GPS L5</td><td>-0.24</td><td>-0.23</td><td>-0.38</td><td>-</td><td>-</td><td>-0.08</td></tr>
<tr><td rowspan="2">GAL E5</td><td>E5a</td><td>-0.30</td><td>-0.19</td><td>-0.29</td><td>-</td><td>-</td><td>-0.08</td></tr>
<tr><td>E5b</td><td>-0.20</td><td>-</td><td>-0.12</td><td>-</td><td>-</td><td>-</td></tr>
</table>

*Entries marked with a '-' indicate that the receiver was not capable of tracking that signal.

## 5. LIVE SATELLITE TESTING

In June 2025, Xona successfully launched its first production-class satellite, Pulsar-0, shown in Figure 10. This IOV satellite features both Pulsar X1 and X5 signals at production power levels. Since entering operation in July of 2025, Pulsar-0 has been demonstrating the compatibility of Pulsar signals with existing GNSS under live-sky conditions.

To monitor the impact of Pulsar on existing GNSS signals, particularly GPS and Galileo for comparison to the analysis shown above, Xona is actively collecting data on a range of COTS receivers from survey-grade to automotive to smartphones during satellite passes. One of the key challenges of live-sky testing is its dynamic nature. The combination of fast motion across the sky and antenna gain pattern makes direct before-and-after comparisons of $C/N_0$ values challenging. Nonetheless, $C/N_0$ can be observed during passes to assess any changes to levels of existing systems.

Figure 11 (a–d) and Figure 12 (a-d) show the $C/N_0$ values of GPS and Galileo signals across four different receivers during a pass of Pulsar-0 operating at nominal received power levels. The tested receivers include the same survey receivers S1 and S2, along with automotive receiver A1 and an Android-based smartphone.

Receivers S1, S2, and A1 were connected to a survey-grade antenna while the smartphone used its native internal antenna, all installed on the roof of Xona's headquarters in Burlingame, California.

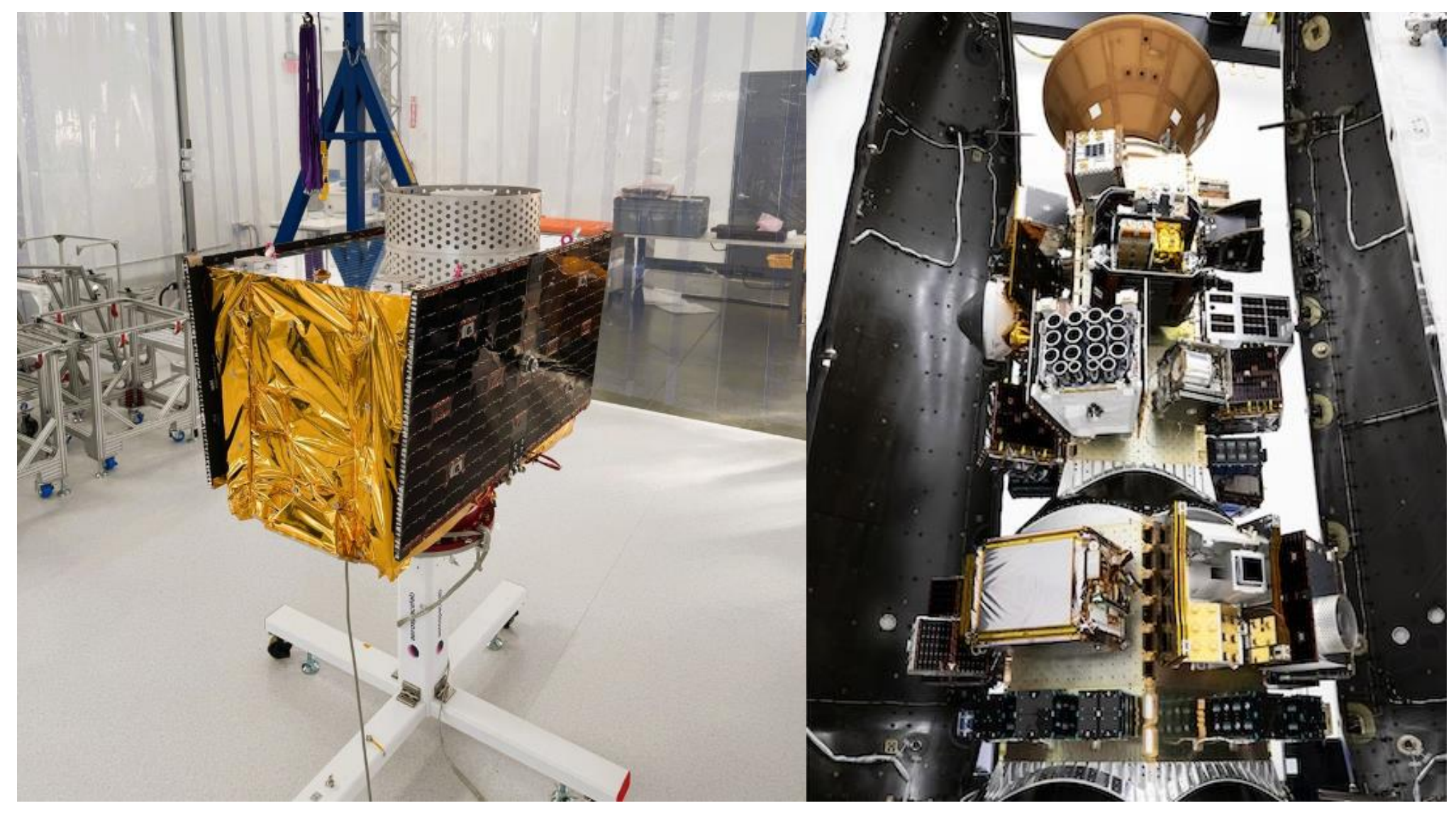

*Figure 10: Pulsar-0 (IOV) satellite in final integration (left) and on the launch vehicle (right).*

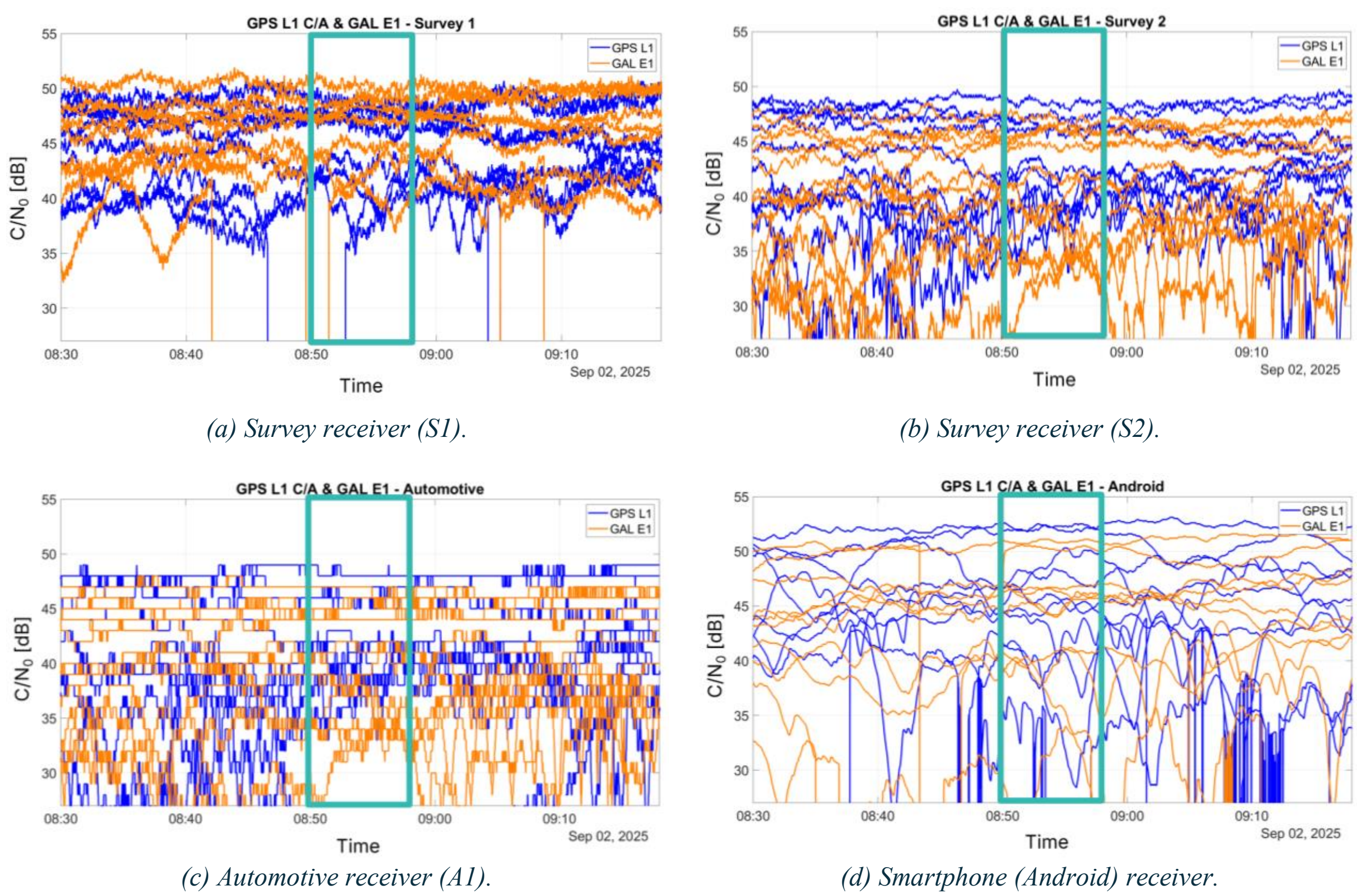

*(a) Survey receiver (S1).*

*(b) Survey receiver (S2).*

*(c) Automotive receiver (A1).*

*(d) Smartphone (Android) receiver.*

*Figure 11: Live-sky C/$N_0$ in L1 / E1 bands during Pulsar-0 satellite passes across several receivers. The turquoise box highlights the pass period of the Pulsar-0 satellite.*

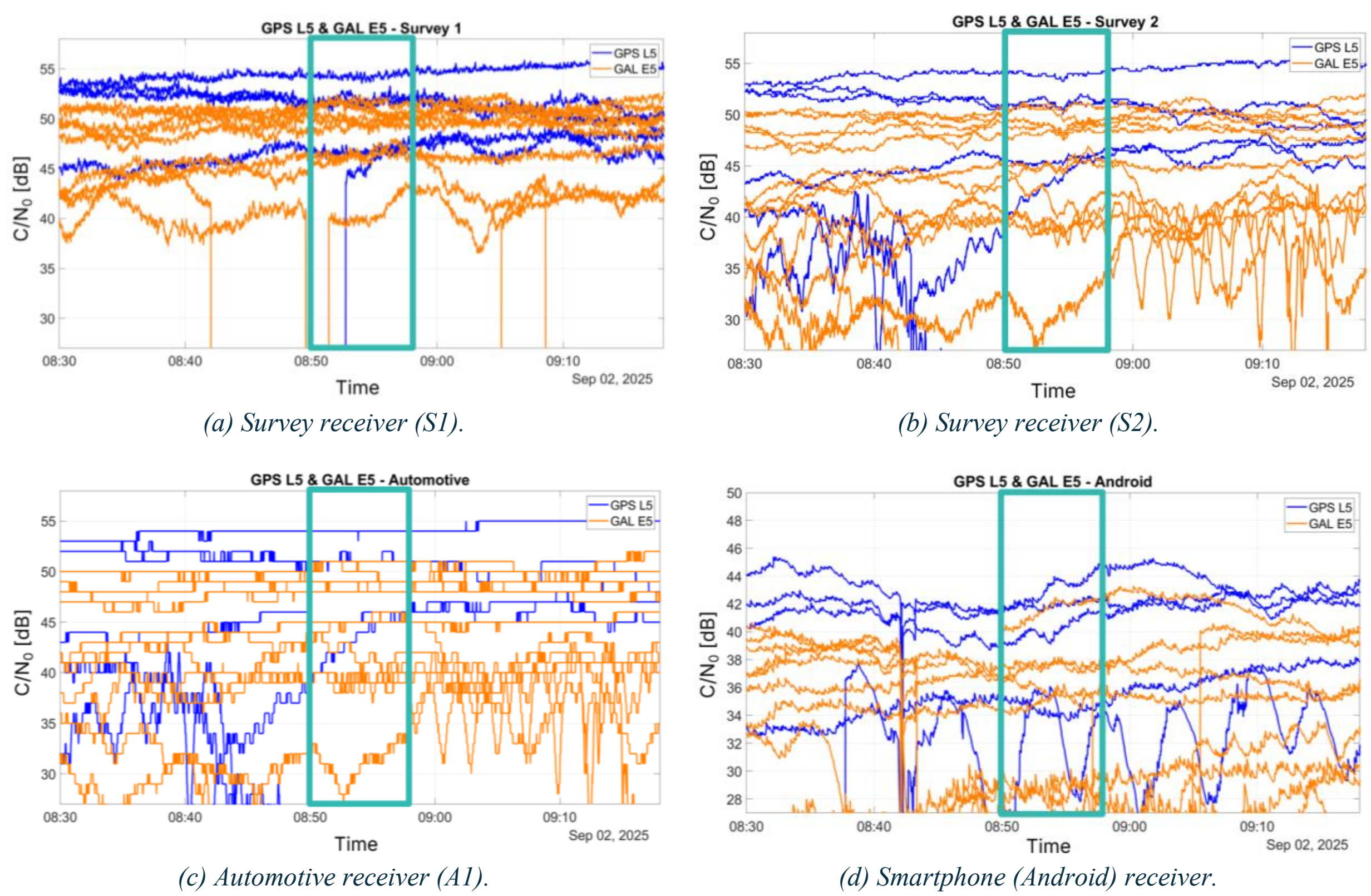

*(a) Survey receiver (S1).*

*(b) Survey receiver (S2).*

*(c) Automotive receiver (A1).*

*(d) Smartphone (Android) receiver.*

*Figure 12: Live-sky $C/N_0$ in L5 / E5 bands during Pulsar-0 satellite passes across several receivers. The turquoise box highlights the pass period of the Pulsar-0 satellite.*

While exact $C/N_0$ degradation values cannot be calculated due to the continuous motion of satellites in the sky, the plots clearly indicate that Pulsar-0 does not cause harmful interference to GPS or Galileo. As expected, the observed $C/N_0$ degradation is lower than theoretical estimates, which assume the presence of the full Pulsar FOC constellation rather than a single satellite. Since the aggregate power from a full constellation depends on each receiver's antenna gain and orientation, actual interference levels can vary across devices. However, these live-sky test results validate that even though the Xona Pulsar signal is approximately 20 dB stronger than GPS and Galileo, it does not result in harmful degradation to current GNSS performance.

## 6. CONCLUSION

This study demonstrates that Xona's Pulsar LEO navigation system is compatible with existing GNSS services in L-band, in particular GPS and Galileo. Through theoretical analysis, hardware testing in the lab, and live satellite trials, Pulsar's X1 and X5 signals have been shown to introduce no meaningful degradation in $C/N_0$ across a variety of commercial GNSS receivers spanning survey-grade to automotive and smartphones. The use of spectrally efficient EFQPSK modulation, combined with careful signal design near GPS L1 and L5 and Galileo E1 and E5, enables coexistence while delivering enhanced performance, including higher power. These results confirm that Pulsar can integrate effectively into the global PNT ecosystem, offering users next-generation capabilities without disrupting the operation of existing GNSS.

## 7. ACKNOWLEDGEMENTS

The authors would like to acknowledge the contributions of Andrew Neish and Eric Lai to early work related to compatibility at Xona.